\documentclass[conference]{IEEEtran}
\usepackage{graphicx}
\usepackage{amssymb}
\usepackage{lineno}
\usepackage{amsmath}
\usepackage{algorithmic}
\usepackage{algorithm}
\usepackage{multirow}
\usepackage{caption}
\usepackage{subcaption}

\usepackage{mathtools}
\DeclarePairedDelimiter\ceil{\lceil}{\rceil}

\usepackage[usenames, dvipsnames]{color}

\def\BibTeX{{\rm B\kern-.05em{\sc i\kern-.025em b}\kern-.08em
    T\kern-.1667em\lower.7ex\hbox{E}\kern-.125emX}}

\begin{document}

\title{A Soft Method for Outliers Detection at the Edge of the Network}

\author{\IEEEauthorblockN{Kostas Kolomvatsos}
\IEEEauthorblockA{Dept. of Computer Science and Telecommunications\\
University of Thessaly\\
Lamia, Greece \\
kostasks@uth.gr}
\and
\IEEEauthorblockN{Christos Anagnostopoulos}
\IEEEauthorblockA{School of Computing Science \\
University of Glasgow\\
Glasgow, UK \\
christos.anagnostopoulos@glasgow.ac.uk}
}

\maketitle

\begin{abstract}
The combination of the Internet of Things and the Edge Computing gives many opportunities to support innovative applications close to end users. Numerous devices present in both infrastructures can collect data upon which various processing activities can be performed. However, the quality of the outcomes may be jeopardized by the presence of outliers. In this paper, we argue on a novel model for outliers detection by elaborating on a `soft' approach. Our mechanism is built upon the concepts of candidate and confirmed outliers. Any data object that deviates from the population is confirmed as an outlier only after the study of its sequence of magnitude values as new data are incorporated into our decision making model. We adopt the combination of a sliding with a landmark window model when a candidate outlier is detected to expand the sequence of data objects taken into consideration. The proposed model is fast and efficient as exposed by our experimental evaluation while a comparative assessment reveals its pros and cons.
\end{abstract}


\section{Introduction}
The current advent of the Internet of Things (IoT) and Edge Computing (EC) opens up the path for the presence of numerous devices around users defining the new form of Pervasive Computing (PC). PC can be seen as the aggregation 
of the two aforementioned vast infrastructures.
The numerous devices are capable of collecting data and performing simple processing activities. They also have the capability of communicating each other as well as with the Cloud back end to transfer data and the produced knowledge. 
The ambient intelligence can be supported by innovative applications
built upon the collected data.
Every node can participate in the envisioned data processing 
that is usually performed upon data streams.
If we consider the nodes present at the EC, we can easily detect the 
ability of executing processing tasks over the formulated geo-distributed datasets reported by a high number of IoT devices.

The aforementioned tasks can demand for simple or more complex processing activities ranging from the delivery of statistical information upon the present datasets (e.g., mean, deviation, median) to 
the conclusion of machine learning models (e.g., regression coefficients, clustering, training of a neural network).
When a request is set, the `baseline' model is to 
launch it across the network and search the information
that end users/applications are interested in \cite{xa}. Obviously, the baseline model 
is prone to an increased network overhead while involving nodes that may not own the appropriate data to efficiently respond to the desired tasks.
A number of research efforts propose techniques for the 
optimal tasks/queries allocation into a number of processing nodes, e.g.,
\cite{dong}, \cite{josilo}, \cite{karanika}, \cite{kolomvatsosapplied}, \cite{kolomvatsoscomputing}, \cite{kolomvatsosfgcs}.
Another challenge is to secure the 
accuracy and the consistency of datasets
at high levels in order to avoid jeopardizing
the quality of the outcomes.
Accuracy is one of the metrics that 
depict the 
quality of data \cite{refcong}.
Accuracy can be jeopardized by the presence of outliers as those objects do not match to the 
remaining objects present in the dataset.
Outliers may lead to heavy fluctuations in the 
underlying data with clear negative consequences 
in the results of any processing activity. For instance, outliers can intensively affect the mean of data which is a critical statistical metric for many tasks.

In this paper, we study a novel model for outliers detection covering a gap in the respective literature. The vast majority of the relevant efforts in the domain adopt a `one-shot' decision making, i.e., when an outlier is detected, the decision is confirmed and final. No additional processing is adopted in the subsequent steps of the decision making. We focus on a mechanism that applies tolerance in the detection process being mainly oriented to support
data streams applications.  
Every outlier data is not directly confirmed as an anomaly in the dataset but we apply 
a temporal management to deliver a set of candidate outliers. Such candidates are confirmed upon the new data that arrive into the system. 
The confirmation of outliers is based on a landmark window expanded to incorporate more data into our process.
We have to notice that, through this approach, we try to avoid scenarios where data objects can be detected as outliers but as new data arrive this decision could be faulty. For instance, at the beginning of a new pattern, the initial data in the new sequence may be seen as outliers but this is not valid in the upcoming steps. 
The proposed approach is simple and adopts
time series management techniques in order to 
detect the patterns in the `behaviour' of data objects.
We elaborate on the detection of change points in the discussed time series and support our mechanism with the appropriate formulations.
For speeding up the envisioned processing to able to support real time applications, we rely on non parametric methods that are capable of exposing the statistics of data objects
in the minimum time and do not require a training phase. 
The following list reports on the contributions of our paper:
\textbf{(i)} we elaborate on a monitoring mechanism for outliers detection upon multivariate data streams;
\textbf{(ii)} we define the concepts of `candidate' and `confirmed' outliers based on a landmark window;
\textbf{(iii)} we provide a set of theoretical models and heuristics to detect outliers; \textbf{(iv)} we adopt the concept 
of the temporal management of data objects for detecting outliers (a data object may be an outlier for a specific window but not if more data arrive). Hence, we do not exclude data from the upcoming processing especially in cases where new data patterns are identified;
\textbf{(v)} we experimentally evaluate the proposed model and compare it with other models found in the literature.
  
The paper is organized as follows.
We report on the related work in Section \ref{related} and present the 
basic information around our problem in Section \ref{preliminaries}.
The description of the proposed approach is performed 
in Section \ref{outliers}.
In Section \ref{evaluation}, we discuss the outcomes of the 
adopted experimental evaluation and conclude this paper in  
Section \ref{conclusions} by giving some of the envisioned 
future research plans.

\section{Related Work}
\label{related}
The detection of outliers is a significant research subject as 
their presence in data may jeopardize the quality of any processing activity upon them. 
The target of any outlier detection method 
is to identify data objects that deviate 
from the distribution of the 
group.
Such objects dictate an `abnormal' behaviour compared to 
the majority of objects in the dataset.
Outliers can be identified upon univariate or multivariate 
data 
and can be easily supported by statistical methods \cite{filzmoser}. 
For instance, the Mahalanobis distance can offer
a statistical view on the correlation of multivariate vectors being based on  
their covariance matrix. 
Such correlations can be also adopted to 
impute data in combination with outliers detection 
for creating efficient models that manage distributed 
data streams \cite{fountas2}, \cite{fountas3}.
Other statistical measures can be found 
in the Cook's distance \cite{cook}, the leverage model \cite{leverage}, 
the $\chi^{2}$ metric 
(it detects deviations from the 
multidimensional normality) and an extended version of the Mahalanobis distance \cite{leys}.
In \cite{baxter}, the interested reader can find a comparison of outlier detection methods.

The authors of \cite{ramasway} identify the top-k
objects that have the highest distance from the population, i.e.,
to their k nearest neighbours.
In \cite{angulli}, the proposed model relies on the average distance
to the k nearest neighbours of each data object. 
In \cite{kontaki}, the authors extend the work presented in \cite{angulli1} and 
focus on multi-query and micro-cluster distance-based outlier detection. 
In \cite{tang}, the authors propose the Relative Density-based Outlier Score (RDOS) algorithm for outliers detection based
on the density of objects.
For estimating this density, the paper 
adopts the Kernel Density Estimation (KDE) method.
The authors of 
\cite{wu} discuss a
semi-supervised model for outliers detection.
focusing on 
data streams. 
In 
\cite{na}, an algorithm for the implicit 
outliers detection is proposed.
The technique relies on the the density of the 
data objects and a distance approximation methodology
to limit the time required to deliver the final responses.
The authors of \cite{huang} focus on a clustering 
approach, i.e., they elaborate on the size of a cluster of outliers that is significantly smaller than other clusters of `normal' data objects.
In \cite{erfani}, the authors study the 
combination of deep belief networks and one-class support vector machine (1SVM). This results a hybrid model 
that manages to identify outliers 
in 
high-dimensional large-scale unlabeled datasets.

In \cite{shaikh}, the authors propose a model for handling uncertainty in the management of outliers 
adopting a probabilistic approach. 
The problem is formulated as 
a top-k distance-based outlier detection upon uncertain data objects. 
Another probabilistic approach is proposed
by \cite{aggarwal}. The target is to 
calculate the probability of having a data object 
in a sub-space located in a region which has a density 
over a pre-defined threshold.
In \cite{cai}, the interested reader can find 
an uncertainty management model 
that proposes a maximal-frequent-pattern-based outlier detection method.
In the respective literature, we can also find
the adoption of  
Fuzzy logic which is the appropriate theory 
to manage the uncertainty present in the discussed problem.
One relevant effort deals with fuzzy regression models
\cite{gal} that try to detect the fuzzy dependencies of
data objects.
In
\cite{cateni}, the interested reader can find
a fuzzy inference system for outliers detection.
This system is compared 
against a statistical approach 
to reveal the pros and cons.
An hybrid model combining Fuzzy Logic and Neural Networks
is described in 
\cite{upasani}.
The paper exposes a Fuzzy min-max neural network adopted to identify outliers in a dataset. 

An outliers detection method can be adopted 
by a system that targets to maintain the accuracy of datasets at high levels. 
With the term accuracy, we denote the 
minimum deviation of data around the mean and limited
fluctuations \cite{kolomvatsosspringer}.
Outliers should be detected when we focus on a 
distributed system where multiple nodes cooperate to execute tasks. 
Usually, nodes exchange 
data synopses for informing their peers about the local datasets \cite{fountas1}, \cite{kolomvatsosdss}, \cite{kolomvatsostkde}.
Data migration or replication may be demanded to formulate the datasets upon which the envisioned processing will be performed. 
The data allocation problem is critical if we want to achieve a fast solution to administrate data streams. 
Data allocation methods should be combined with outliers detection models to secure the minimum acceptable data quality before any processing takes place \cite{kolomvatsosscalcom}.
Data replication mainly deals with the minimization of the latency in the provision of responses and the support
of fault tolerant systems.
Any replication action targets to have data stored at multiple locations in order to avoid 
migration that is negatively affecting the network
overhead.
Data replication is a technique usually utilized 
in Wireless Sensor Networks (WSNs) \cite{kumar}.
The significant is that any outliers detection scheme should take into consideration the distributed nature of data being collected by different sources.
In any case, one can perform 
a selective 
replication under constraints to avoid 
the transfer of high volumes of data in the network
\cite{tai}.
Outliers detection and data replication should be carefully designed when 
data hosts 
are characterized by limited resources (e.g., energy) \cite{boru}. 
In any case, both techniques are part of the
pre-processing phase where data are prepared to be the subject of further processing
\cite{cappiello}.

\section{Preliminaries \& High Level Description of the Proposed Scheme}
\label{preliminaries}
Our setup involves a set of processing nodes (e.g., EC nodes) 
$\mathcal{N} = \left\lbrace n_{1}, n_{2}, \ldots, n_{|\mathcal{N}|} \right\rbrace$
that are the owners of distributed multivariate datasets.
In these datasets, a number of vectors are stored, i.e., 
$\mathbf{x} = \left\langle x_{1}, x_{2}, \ldots, x_{M}\right\rangle$ 
($M$ is the number of dimensions).
Data vectors $\mathbf{x}$ are reported by devices responsible to 
collect them from their environment (e.g., IoT devices).
Without loss of generality, we assume that 
data vectors arrive at discrete time instances 
$t \in \mathrm{T}$ being stored locally for further processing.
The target is to detect outlier vectors 
based on their distance from the population, i.e.,
the local dataset.
We consider that the detected outliers are evicted from the local dataset 
if they are confirmed as dictated by the proposed model.

We rely on a combination of a sliding window and a landmark window approach. Initially, we identify potential outliers, i.e., 
$\tilde{\mathbf{x}}_{1}, \tilde{\mathbf{x}}_{2}, \ldots$ in the last $W$
observations (sliding window). We nominate this set as the `candidate' outliers annotated 
for further investigation. 
When, in a sliding window, we get a number of candidate outliers, we 
also alter our processing and adopt a landmark window to incorporate more data objects 
into our processing.
The maximum size of the landmark window is $\xi \cdot W$
(e.g., $\xi = 2$).
Based on the landmark window, we are able to 
identify the status of each candidate outlier and conclude 
the confirmed outliers
$\overline{\mathbf{x}}_{1}, \overline{\mathbf{x}}_{2}, \ldots $ that will be evicted by 
the local dataset.
With this approach, we try to be aligned with data seasonality or scenarios where a new pattern is initiated by the streams. 
Hence, we postpone the confirmation of an outlier before it is evicted by the local dataset and, thus, its participation in any
local processing. Some objects may not be outliers after the arrival of additional data. For instance, new data may lead to new patterns or new sub-spaces that were not visible in the
current monitored window.

We rely on a distance based outliers detection technique, i.e., we elaborate on the distance 
of $\mathbf{x}_{t}$ from the population
$\mathcal{D}_{W} = \left\lbrace \mathbf{x}_{t} \right\rbrace, t = 1, 2, \ldots, W$.
It should be noticed that our model can be combined with any outlier detection scheme. For instance, we could rely on 
a statistical model, the nearest neighbours scheme, a machine learning mechanisms and so on and so forth.
The distance from the population
is calculated over the mean of the $k$ highest distances 
defined upon the group of 
data objects.
Let the distance of $\mathcal{x}_{t}$ from 
$\mathcal{D}_{W}$ be $d_{t} = f(\mathbf{x}_{t}, \mathcal{D}_{W})$ where $f(\cdot)$ is a function (described later) that quantifies the difference from the population.
We propose the use of a specific function that gives us the opportunity to adopt a `soft' approach in the delivery of the discussed distance and support our temporal tolerance mechanism.
In the proposed model, we elaborate on the monitoring of $d_{t}$ in the landmark window for 
data objects arriving at 
$t \in [W+1, \xi \cdot W]$.
In the aforementioned interval, two actions are performed: (i) the 
detection of candidate outliers; (ii) the confirmation of every candidate outlier in the 
expanded window.
This is a continuous process that keeps track of 
the sliding as well as the landmark windows.
Outliers may change their status from `candidate' to `confirmed' if the trend of their distance from $\mathcal{D}_{W}$ still remains at high levels or increases over time. 
Otherwise, $\mathbf{x}_{t}$ is considered as 
a `normal' data object and is selected to participate in the local dataset and the envisioned processing activities. 

\section{Outliers Detection based on Landmark Windows} 
\label{outliers}
\textbf{The Magnitude of Outliers}. 
Assume that $\tilde{\mathbf{x}}_{t}$ is detected as a candidate outlier in the sliding window
$[(\omega - 1) \cdot W + 1, \omega \cdot W]$ ($\omega \in \left\lbrace 1, 2, \ldots \right\rbrace$).
We define the concept of the magnitude of the candidate outlier
$\tilde{\mathbf{x}}_{t}$ based on its distance to
$\mathcal{D}_{W}$ as represented by $d_{t}$.
We consider that the `fuzzy' notion 
of the magnitude of each outlier is measured 
by a sigmoid function, i.e.,
$\lambda = \frac{1}{1+e^{\left(-\alpha x + \beta\right)}}$
where $\alpha$ \& $\beta$ are smoothing parameters.
When $d_{t}$ exceeds a threshold (as defined by the realization of 
the aforementioned sigmoid function), the magnitude of the
outlier indication for $\tilde{\mathbf{x}}_{t}$ is very high 
(close to unity).
When $d_{t}$ indicates that $\tilde{\mathbf{x}}_{t}$ is close to the
population, $\lambda$ becomes very low and gets values close to zero.
$\lambda$ is recorded to assist us 
in the confirmation of 
$\tilde{\mathbf{x}}_{t}$ as an outlier while altering the processing 
by moving from the sliding window scheme to the 
landmark window model.

\begin{figure}[h!]
 \centering
 \includegraphics[width=0.5\textwidth]{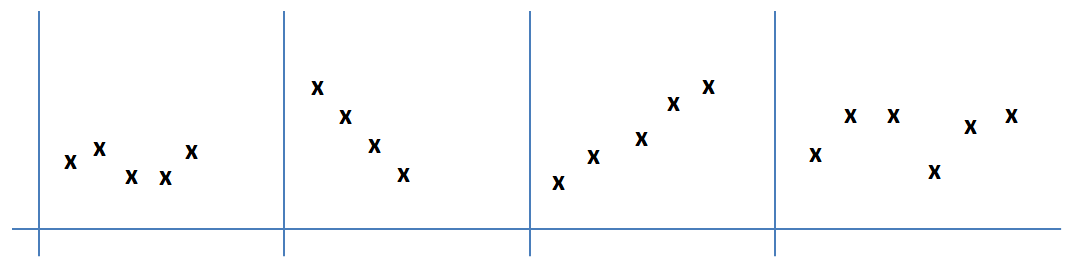}
 \caption{Some example patterns for $\lambda$} 
 \label{fig:lambda}
\end{figure}

When the adoption of the landmark window is decided (this means that we detect the presence of 
candidate outliers in the sliding window),
we increase the size of the window $W$ by adding a small amount of discrete time instances,
i.e., $w = a \cdot t$ with $a$ being a small positive number.
Upon these time instances, we record and monitor the realization of
$\lambda$ for $\tilde{\mathbf{x}}_{t}$, however, 
taking into consideration an increased number of
data objects into our calculations. 
This means that we perform again the calculations for exposing the distance of 
candidate outliers from the population.
The discussed processing is performed in parallel with the monitoring
for detecting new candidate outliers as the sliding window is updated.
Adopting the aforementioned approach,
we generate a time series
of $\lambda$ values for each 
candidate outlier as follows:
$\lambda_{1}, \lambda_{2}, \ldots, \lambda_{y}$
for time steps $W+w, W+2 \cdot w, W+3 \cdot w, \ldots, 
W + y \cdot w$ where $y = \ceil*{\frac{(\xi - 1) \cdot W}{w}}$.
The trend of the magnitude $\lambda$ plays a significant 
role in the final decision on if $\tilde{\mathbf{x}}_{t}$
will be finally annotated as a confirmed outlier or not.
In Figure \ref{fig:lambda}, we can see some example patterns
for $\lambda$. The magnitude can be altered as new data objects arrive and participate into 
our processing.
For instance, assume that our outliers detection scheme is based on a clustering approach.
$\tilde{\mathbf{x}}_{t}$ is considered as the object with a high distance from 
the `normal' clusters. 
As new data objects are observed, we can incrementally update the detected clusters and update the 
corresponding centroids accordingly.
This means that the distance of $\tilde{\mathbf{x}}_{t}$ from the clusters can 
increase if the new data objects are located at the `opposite' side in the data space compared to $\tilde{\mathbf{x}}_{t}$,
decrease if new data objects are located at `same direction' in the data space compared to $\tilde{\mathbf{x}}_{t}$ 
and so on and so forth.
$\tilde{\mathbf{x}}_{t}$ can also represent the beginning of a new cluster and the new data objects may be located around it.
The proposed model tries to detect the trend of the magnitude, then, to 
conclude to the final annotation 
of $\tilde{\mathbf{x}}_{t}$.

\textbf{Trend Analysis \& Magnitude of Outliers}.
As our focus is on a streaming environment, we avoid adopting
parametric methods that require training and rely on fast techniques 
to quickly elaborate on the magnitude of outliers and its trend.
Our non parametric trend analysis is applied 
upon $\lambda$ realizations as the landmark window is expanded
subject to the presence of candidate outliers.
We try to catch the variability in the magnitude of the candidate 
outliers that may be due to many factors like seasonal cycles, variations in the incoming data, natural evolution of new patterns and so on and so forth. In our model, for trend analysis, we adopt 
an ensemble scheme upon the widely known Mann-Kendall metric or Mann-Kendall test (MKM) \cite{kendall}, \cite{mann} and the Sen's slope (SS) \cite{sen}.

The MKM
is adopted to 
indicate if there are trends in a time series sequence, i.e., $\left\lbrace \lambda_{i} \right\rbrace, i=1,2, \ldots, y$.
Our target is to expose the temporal 
variation of the magnitude of a candidate outlier $\tilde{\mathbf{x}}_{t}$. It is a non parametric method which makes it useful to be applied in a streaming environment and limit the need for complex methodologies. Its rationale is located around the idea to perform a statistical processing upon the observed data and not on random variables.
It pays attention on the sign of 
the difference between the observed data and previous measurements and compares every later-measured data with previous observations in pairs.
For instance, if we focus on the $W$ magnitude values, the method requires $\frac{W(W-1)}{2}$ comparisons upon pair of observations. The complexity of the process is $O(W^{2})$; in general $W$ is low compared to the 
total number of the recorded observations, thus, the proposed processing 
can be adopted for streams management.
The significant is that the adopted method 
is not affected by missing values and is not based on any assumption about the distribution of data.
Moreover, the MKM metric is invariant to transformations (e.g., logs) enhancing its applicability in multiple application domains.
The MKM is realized upon the following equation: 
$S = \sum_{i=1}^{W + h \cdot w -1} \sum_{j=i+1}^{W + h \cdot w} sign(\lambda_{j} - \lambda_{i})$
where $h$ is the number of added slots to the landmark window which depicts the total number of $\lambda$ realizations incorporated into the MKM calculations.
Additionally, $sign(\lambda_{j} - \lambda_{i})$ is considered equal to unity if $\lambda_{j} > \lambda_{i}$, equal to zero if $\lambda_{j} = \lambda_{i}$ and equal to -1 when $\lambda_{j} < \lambda_{i}$. Upon $S$, we can define the parameter $Z$ realized as follows:
$Z = \frac{S-1}{var(S)}$ if $S>0$, $Z = 0$ if $S=0$ and $Z = \frac{S+1}{var(S)}$ if $S<0$
where $var(S) = \frac{S\cdot(q-1)\cdot(2q+5)- \sum_{i=1}^{q'} tp\cdot(tp-1)\cdot(2tp+5)}{18}$,
$q = W + k \cdot w$, $tp$ is the ties of the $p$th value and $q'$ is the number of ties.
Finally, if $Z$ is positive, we can conclude an increasing trend and the opposite stands when $Z$ is negative. When testing two sided trends, the null hypothesis of no trend is rejected is $|Z|>Z_{\phi/2}$ with $\phi$ being the significance level. 

The SS is also non parametric method being calculated as the median of all the slopes estimated between sequential data of the time series. The following equation holds true:
$SS = median\left[\frac{\Delta \lambda}{\Delta t} \right]$ where $\Delta \lambda$ is the difference between sequential $\lambda$ realizations and $\Delta t$ depicts the change in time. 
When $SS>0$, we identify an increasing trend and the opposite stands for 
$SS<0$. 

We rely on a simple and fast, however, efficient technique to aggregate $S$ \& $SS$ and support our decision about the trend of the magnitude $\lambda$.
The easy scenario is met when both techniques agree upon the trend of $\lambda$ (increasing or decreasing trend). In case of an disagreement, we consider a `strict' boolean model which relies on a conjunctive form. This means that disagreements are solve by deciding a `neutral' view for $\lambda$. 
If the final outcome indicates an increasing
trend, we update the status of $\tilde{\mathbf{x}}_{t}$ and
retrieve it as a confirmed outlier $\overline{\mathbf{x}}_{t}$ having it rejected from any further processing.
If the tread is detected as decreasing or neutral and 
$d_{t} < \theta$ ($\theta$ is a pre-defined threshold indicating a low distance with the remaining population),
$\tilde{\mathbf{x}}_{t}$ is accepted as a normal value.

\section{Experimental Evaluation}
\label{evaluation}
\textbf{Simulation Setup and Performance Metrics}.
We report on the experimental evaluation of the proposed model 
based on a custom simulator built in Java (an individual Java class
is adopted to realize our simulator). 
The simulator performs the processing of two real datasets 
where a number of outliers are detected and applies the proposed approach 
to reveal if it is capable of identifying the 
reported outliers.
In our experiments, we rely on the following datasets (they depict multivariate data) \cite{keller} :
\textbf{(i) Dataset 1}. The ionosphere dataset has 32 numeric dimensions and 351 instances where 126 outliers (35.9\%) are detected. Inliers are good radar signals showing evidence of some kind of structure in the ionosphere while outliers are bad radar data for which signals pass through the radar;
\textbf{(ii) Dataset 2.} The Wisconsin Prognostic Breast Cancer (WPBC) dataset has 33 numerical dimensions and 198 instances where 47 outliers (23.74\%) are detected.
Our evaluation targets to show if the proposed model is capable 
to identify the already reported 
outliers adopting the sliding/landmark window models. Actually, our results 
will reveal how many already identified outliers are confirmed as new data 
are taken into consideration. We have to notice that, compared to other `legacy' outlier detection methods, the proposed approach does deal with the entire dataset (the total number of vectors)in the envisioned calculations but with the aforementioned 
sliding/landmark window approaches.
The performance is evaluated upon a set of metrics dealing with the accuracy of the detection, the precision, the recall and the curve known as the Receiver Operating Characteristic (ROC)
\cite{campos}.
Accuracy $\epsilon$ is defined as the number of correct detections out of the 
total number of the identified outliers into the above described datasets, i.e.,
$\epsilon = \frac{|O_{d}|}{|O|}$ where $O_{d}$ is the set of the detected outliers as decided 
by our model and $|O|$ is the set of the reported outliers in the above described datasets.
Precision is defined as the 
the fraction of the correctly detected outliers, i.e., $v = \frac{TP}{TP+FP}$ where 
$TP$ are the correctly detected outliers and $FP$ are objects that should not be detected
as outliers.
Recall is defined as the fraction of the detected outliers that are successfully retrieved compared to the
true outliers present in the dataset, i.e., $r=\frac{TP}{TP+FN}$ where $FN$ is the number of 
outlier objects that are not detected by our model.
The F-measure is a combination of $v$ and $r$ defined as follows:
$\phi = 2\cdot \frac{v \cdot r}{v + r}$.
Finally, the ROC curve is obtained by plotting all possible combinations of 
true positive rates and false positive rates. The curve can be depicted 
by a single value known as the area under the ROC curve (ROC AUC) \cite{campos}. 
This value can be seen as the mean 
of the recall upon the top-ranked objects (in the list of the potential outliers). 
We have to notice that, in our model, we adopt 
a `binary' scheme, i.e., an object is an outlier or not, thus, the score for each object is 
realized to unity or zero. 
In our simulations, we consider $W \in \left\lbrace 5, 10 \right\rbrace$,
$W \in \left\lbrace 2, 3, 4, 5 \right\rbrace$, $\alpha = \beta = 2$, $w=3$ and
$\xi = 2.5$.

\begin{figure}[h!]
 \centering
 \includegraphics[width=0.55\textwidth]{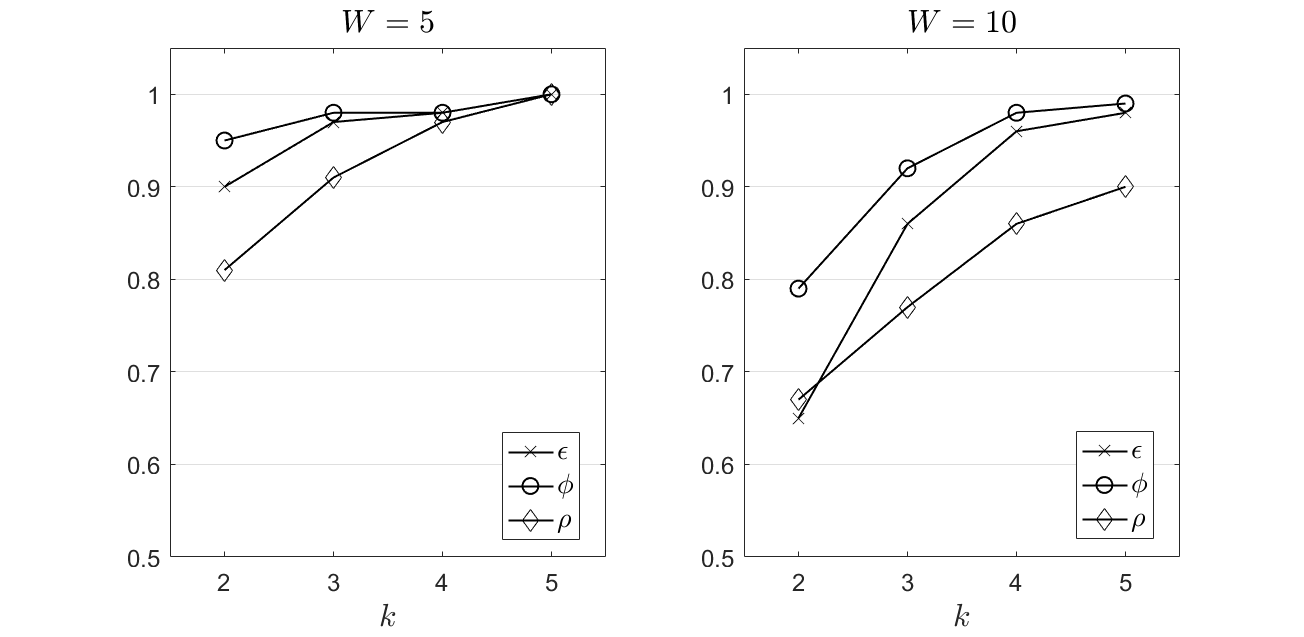}
 \caption{Performance outcomes for Dataset 1} 
 \label{fig:fig1}
\end{figure}

\textbf{Experimental Evaluation}.
We report on the performance of the proposed system
concerning the above described metrics $\epsilon$, $\phi$ \& $\rho$.
In Figure \ref{fig:fig1}, we present our results 
for the Dataset 1. 
We observe that an increased number of neighbours taken 
into consideration to depict the distance of a data object 
with the population (the $k$ highest distances) positively affects the 
performance of our model.
As $k$ increases, the adopted metrics reach very close to the 
optimal value.
In particular, the accuracy $\epsilon$ stays close to unity
exhibiting the ability of the proposed solution to successfully detect the 
reported outliers. 
The same observation stands for $\phi$ which depicts the capability of our scheme to keep 
the precision and the recall close to the 
maximum value, thus, false positives and 
false negatives events are minimized.
The area under the ROC also is concluded at high values
which represents the optimality in the delivery of 
the true positive and false positive rates.
All the above discussion refers in the experimental scenario where $W=5$. Obviously, a limited
number of values taken into consideration in our calculations in combination with the landmark window approach leads to the best performance.
When $W=10$ (see Figure \ref{fig:fig1} - right), we observe a similar performance, however, the outcomes are lower than in the previously presented experimental scenario (especially for $k leq 4$ and $\phi$, $\rho$ metrics).
This outcome is due to the 
lower number of confirmed outliers by our model compared to the case where $W=5$. Recall that the sliding window $W$ is expanded when we switch to the landmark window model and more data are taken into consideration into the envisioned calculations.
Hence, a sub-set of candidate outliers are not finally confirmed and are incorporated into the dataset. 
This aspect reveals the new approach that our model proposes in the outlier detection domain. The discussed values can be part of the upcoming processing as their distance from the population, for the specific window, does not excuse their eviction from the local dataset.
We have to remind that our decision making does not consider the entire dataset to detect outliers, thus, an object may not be an outlier based on data seasonality but it could be an outlier if considered compared to entire stream.
If we focus on other models and the analysis described in 
\cite{campos} for the same dataset, we can detect that 
other efforts achieve the maximum $\phi$ in the interval [0.75, 0.88] while
$\rho$ is realized in the interval [0.82, 0.96]
\footnote{www.dbs.ifi.lmu.de/research/outlier-evaluation/DAMI/literature/Ionosphere/}. 
We observe that the proposed model outperforms other relevant mechanisms especially for an increased $k$.

\begin{figure}[h!]
 \centering
 \includegraphics[width=0.55\textwidth]{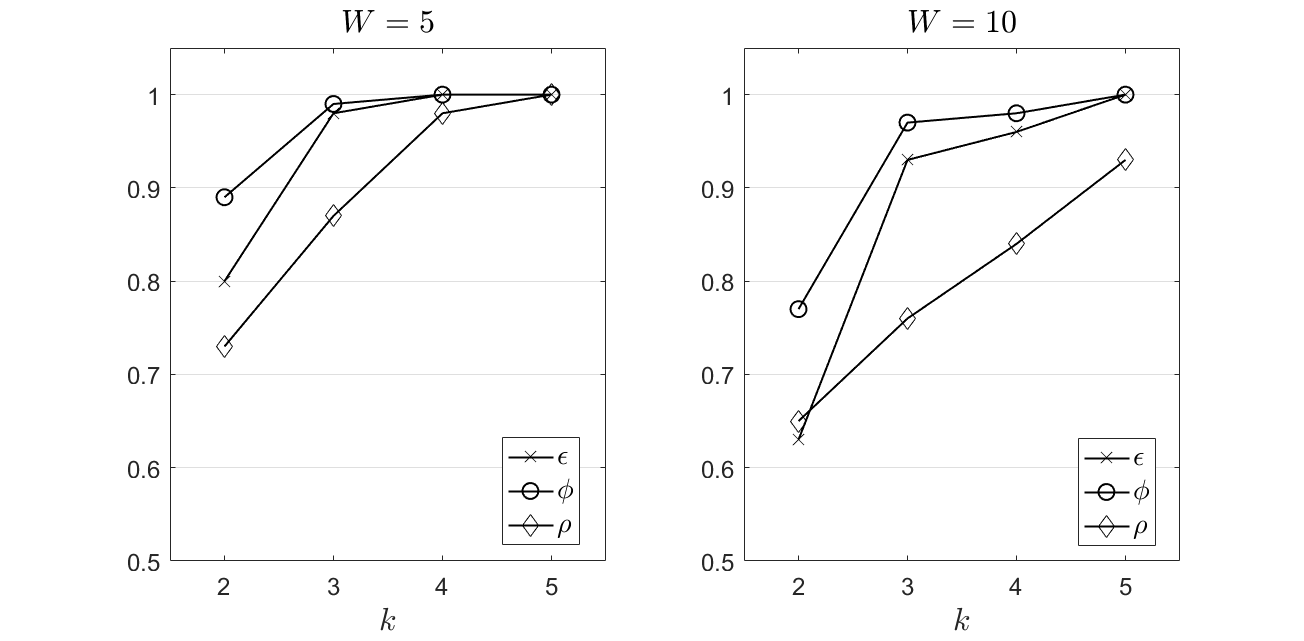}
 \caption{Performance outcomes for Dataset 2} 
 \label{fig:fig2}
\end{figure}

In Figure \ref{fig:fig2}, we present our results for the 
Dataset 2. We observe a similar performance as in the experimentation upon the Dataset 1 with slightly lower results for the envisioned metrics. In any case, the discussed evaluation outcomes confirm our previously presented observations. 
Other relevant models, evaluated in
\cite{campos} for the same dataset, achieve 
the maximum $\phi$ in the interval [0.38, 0.44] while
$\rho$ is realized in the interval [0.46, 0.58]
\footnote{www.dbs.ifi.lmu.de/research/outlier-evaluation/DAMI/literature/WPBC/}. 
We observe that the proposed model clearly outperforms these efforts for various realizations of $k$.

We also focus on the time required to perform the proposed calculations
and proceed with the confirmation of outliers.
For the Dataset 1, the proposed model needs (in average)
0.097 and 0.287 ms for $W=5$ and $W=10$, respectively.
For the Dataset 2, our model requires (in average)
0.102 and 0.277 ms for $W=5$ and $W=10$, respectively.
We easily discern the ability of our model to support real time 
applications as the throughput reaches 
the management of 
[3484, 10309] multivariate data objects per second (approximately).
This is very significant when we focus on very dynamic environments where it is imperative to conclude the final decision making in limited time.

\section{Conclusions \& Future Work}
\label{conclusions}
We propose the use of a model that, based on a `soft' approach, decides the presence 
of outliers in a dataset. We focus on streaming environments and 
a sequential scheme for delivering the final decision.
We define the concepts of candidate and confirmed outliers as well as 
the magnitude of the difference of an outlier from the remaining population.
Our temporal management process builds upon the combination 
of a sliding with a landmark window to expand the observations taken into consideration before we conclude a confirmed outlier.
The proposed technique is experimentally evaluated 
and its advantages and disadvantages are revealed. 
Its speed for delivering the final outcome is due to the adoption
of non parametric models that can derive the trend of the magnitude of potential outliers in the minimum time.
The comparative assessment positions our technique in the respective literature.
Our future plans involve the adoption of a scheme based on Fuzzy Logic 
and machine learning to be able to expose more complex trends and 
connections between data objects.

\end{document}